*Article*

# Human-like Energy Management Based on Deep Reinforcement Learning and Historical Driving Experiences

**Hao Chen**[a, b], **Xiaolin Tang**[a, *], **Guo Hu**[b], **Teng Liu**[c, a]

*a*. Department of Automotive Engineering, Chongqing University, Chongqing 400044, China;

*b*. Department of Industrial Design, Chongqing Normal University, Chongqing 401331, China;

*c*. Clinical Research Center and Medical Pathology Center (MPC), Chongqing University Three Gorges Hospital, Chongqing University, Wanzhou, Chongqing, P.R. China.

E-Mails: cquchenhao@163.com (H.C.); tangxl0923@cqu.edu.cn (X.T.); 38515820@qq.com (G.H.); tengliu17@gmail.com (T.L.). *Corresponding author.

---

**Abstract:** The development of hybrid electric vehicles relies on advanced and efficient energy management strategies. With online and real-time requirements in mind, this paper introduces a human-like energy management framework for hybrid electric vehicles based on deep reinforcement learning methods and collected historical driving data. The studied hybrid powertrain features a series-parallel topology, and a control-oriented model for the hybrid powertrain is established initially. Subsequently, a unique deep reinforcement learning algorithm, the deep deterministic policy gradient (DDPG), is presented. To enhance the power split control derived within the deep reinforcement learning (DRL) framework, globally optimal control trajectories obtained from dynamic programming (DP) are utilized as expert knowledge to train the DDPG model. This approach serves to augment the optimality of the proposed control architecture. Furthermore, historical driving data gathered from experienced drivers are employed to replace DP-based controls, thus constructing human-like EMSs. Lastly, diverse categories of experiments are conducted to assess the optimality and adaptability of the proposed human-like EMS. The improvements in fuel economy and convergence rate indicate the efficacy of the constructed control structure.

---

## 1. Introduction

Hybrid Electric Vehicles (HEVs) and plug-in Hybrid Electric Vehicles (PHEVs) hold significant potential for energy conservation and emission reduction, positioning them as leading contenders within the contemporary automobile sales market [1-5]. Enhancing fuel economy and upholding optimal performance for these vehicles are contingent upon two pivotal technologies: energy management and powertrain matching [6,7]. Presently, devising an optimal power allocation strategy for HEVs presents a challenging endeavor [8-10].

The intricate integration of multiple energy storage resources within hybrid powertrains necessitates a sophisticated Energy Management System (EMS) to facilitate seamless power coordination across diverse driving scenarios [11, 12]. Thus far, three distinct categories of EMSs have been introduced to address energy management challenges, particularly power split dilemmas, across various powertrain architectures. They are rule-based, optimization-based, and learning-based policies [13-15]. However, the translation of these commendable EMS strategies into practical real-world driving contexts to establish human-like power split controls, which remains a prevailing research imperative within the contemporary energy management domain.

Fueled by the remarkable advancements in deep learning and reinforcement learning (RL) within the field of artificial intelligence, deep reinforcement learning (DRL) has emerged as a promising methodology for developing intelligent EMS tailored for hybrid powertrains [16, 17]. As an illustration, a prominent instance in [18] showcased the application of well-established deep Q-learning (DQL) to address the continuous optimization control challenge within energy management, resulting in impressive performance outcomes. To enhance convergence rate, Qi et al. [19] employed dueling Deep Q-Network (DQN) to tackle the energy management problem specific to a parallel powertrain, validating the superiority of the proposed control policy over conventional onboard binary control. Additionally, References [20] and [21] harnessed deep deterministic policy gradient (DDPG) to derive optimal EMS for the Prius and series-parallel PHEV, respectively. The resultant DDPG-driven control policies were benchmarked against other DRL methods, affirming their superior fuel economy. Nonetheless, the presence of multiple neural networks (NN) in these DRL approaches renders the process of obtaining adaptable DRL-based EMS time-intensive. Consequently, these derived control policies are not readily deployable within real-world driving contexts.

This article presents a human-like energy management framework utilizing the DRL technique in conjunction with gathered historical driving data, as depicted in Fig. 1. Initially, a series-parallel hybrid powertrain is meticulously established as the testified target of the proposed EMS. Subsequently, a refined approach to constructing DDPG-based controls is introduced, where the expertise provided by dynamic programming (DP)-derived optimal global controls is leveraged to expand the search space for control actions. These control strategies are informed by historical driving data accumulated from seasoned drivers. This process yields a human-like driving policy with assured optimality and convergence rate. Lastly, both standard and real-world driving cycles are employed to assess the effectiveness, optimality, and adaptability of the newly proposed human-like EMS.

This paper encompasses three distinctive contributions and innovations: 1) Introducing a human-like EMS grounded in the DRL methodology and informed by collated historical driving data; 2) Establishing an enhanced DDPG framework that integrates DP-informed optimal control actions; 3)

Harnessing real-world driving behaviors as a guiding force, enabling the proposed human-like control policy to seamlessly adapt across varying driving cycles. The amalgamation of the DRL technique and real-world driving data presented, and herein offers a potential solution for online or real-time power split controls for HEVs and PHEVs.

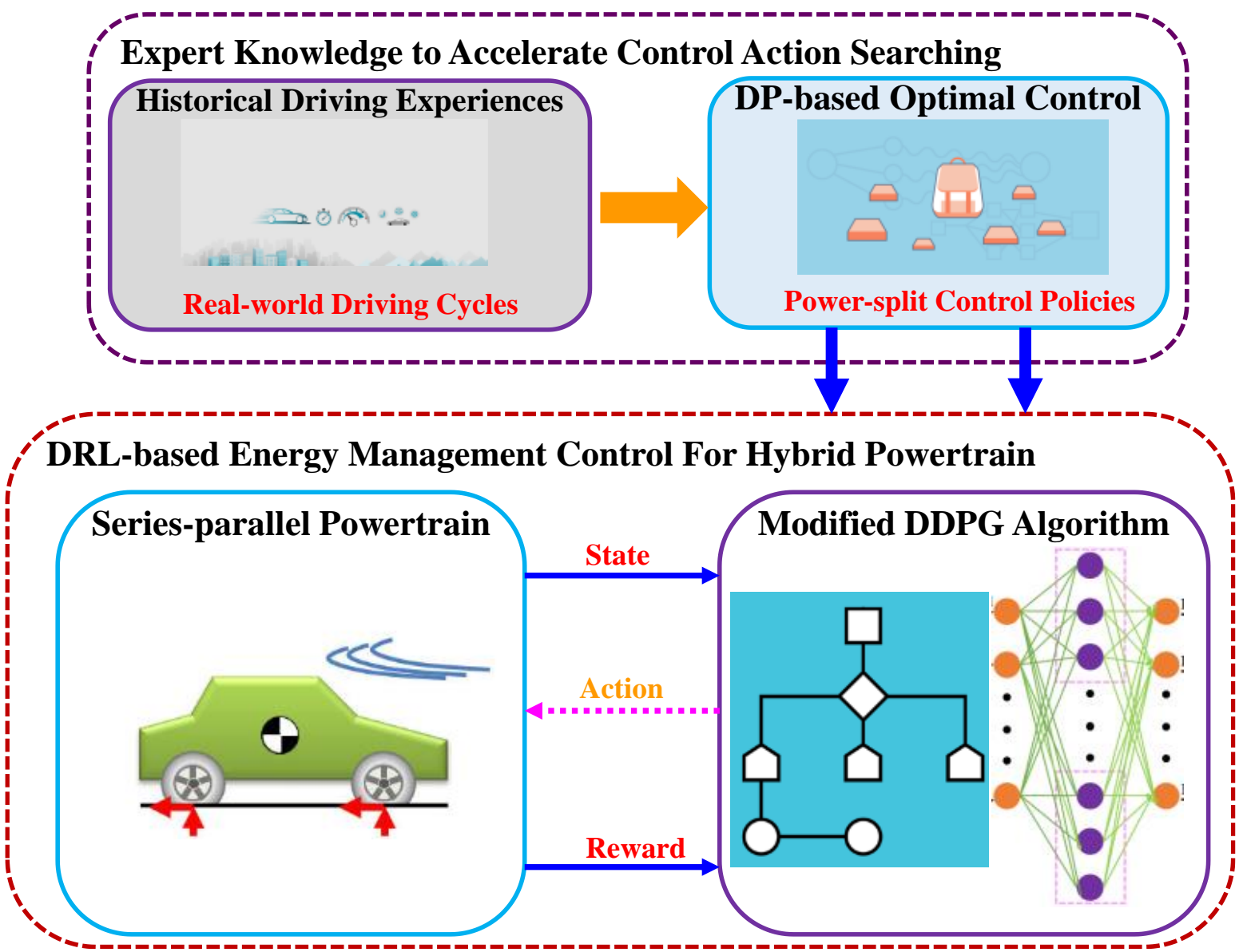

**Figure 1.** Human-like energy management system with expert knowledge for HEV.

The following construction of this paper is organized as follows: Section II presents the series-parallel hybrid powertrain along with the pertinent energy management challenge. In Section III, we introduce the enhanced DDPG technique, the DP method, and the incorporation of real-world driving data. Section IV outlines the simulation results used to evaluate the proposed control structure, analyzing and verifying its optimality and adaptability. Lastly, Section V offers a summary of the conclusion and future research directions.

## 2. Powertrain Modeling and Energy Management Problem

The hybrid powertrain under investigation exhibits a series-parallel topology, and its control architecture is illustrated in Fig. 2 [22]. Its key components encompass the internal-combustion engine (ICE), lithium-ion battery pack, traction motor, and integrated starter generator (ISG). The energy management controller facilitates the allocation of output power between the ICE and battery to achieve optimization control objectives. This section expounds upon the mathematical rigor of this powertrain model, while Table I presents the values of crucial parameters.

### *2.1. Powertrain Modeling*

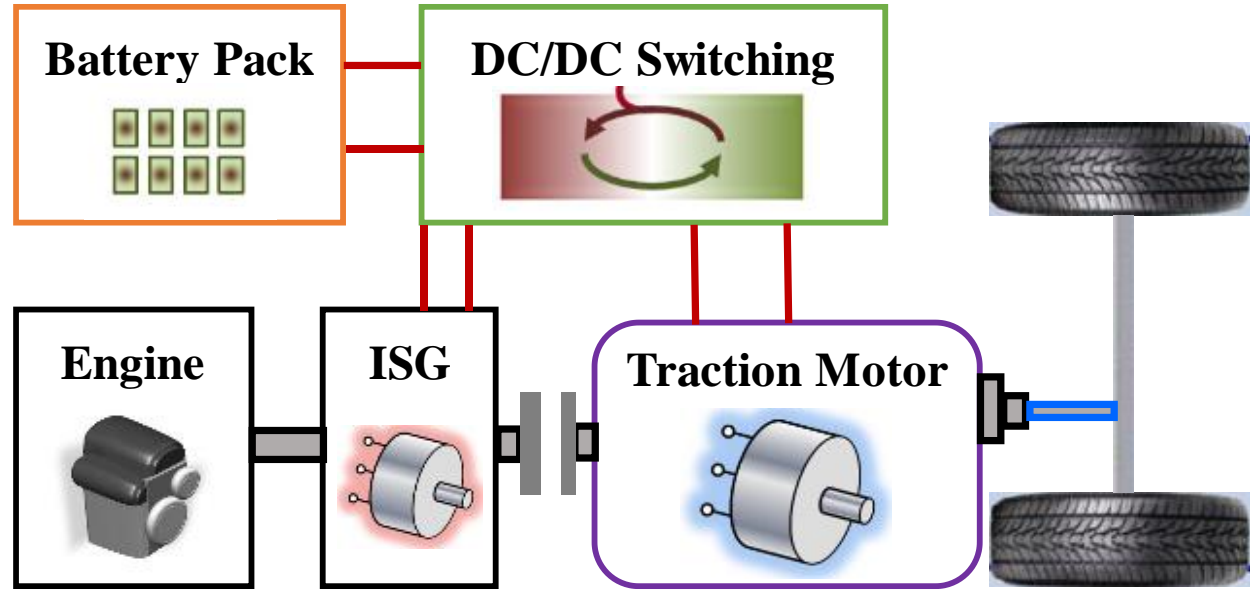

**Figure 2.** Configuration of studied series-parallel powertrain topology [22].

With the driving cycle predetermined, the vehicle's speed and acceleration become established. Consequently, the power demand ($P_d$) for the entire powertrain is delineated into three components, as detailed below:

$$P_d = P_r + P_a + P_i \tag{1}$$

$$P_r = f \cdot M_v \cdot g \cdot v \tag{2}$$

$$P_a = \frac{1}{2}\rho A_a C_D v^2 \cdot v \tag{3}$$

$$P_i = M_v a \cdot v \tag{4}$$

where $P_r$, $P_a$, and $P_i$ are the powers related to the rolling resistance, aerodynamic drag, and inertial force, respectively. $M_v$ is the curb weight, $g$ is the gravity coefficient, $\rho$ is the air density, $A_a$ is the frontal area, $f$ and $C_D$ are the coefficients of rolling resistance and aerodynamic drag, respectively. $v$ and $a$ are the vehicle speed and acceleration and they are mutable in different driving cycles, and hence the power demand changes with the driving cycle. This underscores the necessity for the energy management controller to dynamically calibrate its EMS in response to diverse driving conditions.

The power requirement is met by two onboard energy storage systems: the battery and the ICE. Within the battery, the State of Charge (SoC) signifies the remaining electric capacity subsequent to completing the designated driving cycle. SoC is quantified on a scale from 0 to 1, and its fluctuation is calculable using the following formula:

$$\dot{SoC} = -I_b / Q_c \tag{5}$$

where $Q_c$ and $I_b$ are the nominal capacity and output current of the battery, respectively. To simulate the battery as internal resistance model [23], the output power $P_b$ and voltage $U_b$ of battery are written as follow:

$$P_b = U_b \cdot I_b \tag{6}$$

$$U_b = V_{oc} - I_b r_0 \tag{7}$$

where $V_{oc}$ is the open-circuit voltage, and $r_0$ implies the internal resistance. Incorporating the Eq. (5) to (7), the variation of SoC is able to be described as:

$$\dot{SoC} = -(V_{oc} - \sqrt{V_{oc}^2 - 4r_0 P_b}) / (2Q_c r_0) \tag{8}$$

Preserving the battery's extended operational lifespan necessitates preventing both overcharging and over-discharging. Consequently, within the scope of this study, the SoC is constrained within the range [0.2, 0.9]. For the examined series-parallel powertrain, the battery boasts a nominal capacity of 8.1Ah, accompanied by a nominal voltage of 200V and internal resistance measuring 0.25Ω. Given that the battery's output power is contingent upon both the power demand and the power supplied by the ICE, the SoC is derivable using Equation (8) at any given time. Thus, SoC is selected as a state variable to gauge the performance of control actions.

The significant parameter in ICE is the fuel consumption rate, which reflects the fuel economy of HEV directly. Modeled by static map method, the fuel consumption rate $m_f$ is determined by the speed and torque of ICE as follow:

$$m_f = f_e(T_e, \omega_e) \tag{9}$$

where $T_e$ and $\omega_e$ are the torque and speed of ICE, respectively. $f_e$ is always represented by the look-up table function, which means the brake specific fuel consumption (BSFC) curve of ICE is mutable in this work.

**Table 1.** [22] Powertrain Parameters for the Studied Series-Parallel HEV

| Symbol | Implication | Values |
|---|---|---|
| $M_v$ | Curb weight | 1325 kg |
| $\rho$ | Air density | 1.225 kg/m$^3$ |
| $f$ | Rolling resistance coefficient | 0.012 |
| $A_a$ | Frontal area | 2.16 m$^2$ |
| $C_D$ | Aerodynamic drag coefficient | 0.26 |
| $g$ | Gravity coefficient | 9.8 m/s$^2$ |
| $V_{oc}$ | Open circuit voltage | 150 V |
| $SoC_{ref}$ | Charge sustaining value | 0.6 |

Within this series-parallel powertrain configuration, the operational speed of the ICE spans from 1000 to 4500 rpm. At its zenith, the ICE yields a peak power output of 57 kW at 5000 rpm, accompanied by a peak torque of 115 Nm achieved at 4200 rpm. Given a predetermined power demand, the ICE's output power dictates both battery power and SoC. Consequently, this article elects to employ ICE power as the designated control actions within a continuous space [22].

### *2.2. Energy Management Formulation*

The energy management problem of HEV is converted into an optimization control problem with a predefined objective and several constraints. The goal of this problem is to search a control sequence to achieve the best control performance. The control objective is usually represented by the cost function $J$ over a finite time horizon as follow:

$$\begin{cases} J = \int_0^T [m_f(t) + \delta(\Delta_{SoC}(t))^2]dt \\ \Delta_{SoC}(t) = \begin{cases} SoC(t) - SoC_{ref} & SoC(t) < SoC_{ref} \\ 0 & SoC(t) \geq SoC_{ref} \end{cases} \end{cases} \tag{10}$$

wherein the first term is the fuel economy. the second one means the charge sustaining restraint. $\delta$ is a positive weighting parameter to tune these two goals in the cost function. It equals 500 in this work. SoC$_{ref}$ is a pre-defined factor (actual experience) to enable the final value of SoC close to its initial value. It is settled as 0.7 in this paper.

Generally, the cost function is affected by the state variable $s \in \boldsymbol{S}$ and $ac \in \boldsymbol{A}$. In this paper, the state variables are the vehicle speed, acceleration and SoC, and the control action is the power of ICE:

$$S=\{v, ac, SoC \in [0.2, 0.9]\} \tag{11}$$

$$A=\{P_e \in [0, 57]\} \tag{12}$$

To choose the best control actions from a normal working area, the defined optimization control problem should follow a couple of constraints. It implies that the ICE, battery, ISG, and traction motor need work in a reasonable range as:

$$\begin{cases} SoC_{\min} \le SoC(t) \le SoC_{\max} \\ P_{b,\min} \le P_b(t) \le P_{b,\max} \end{cases} \tag{13}$$

$$\begin{cases} \omega_{x,\min} \le \omega_x(t) \le \omega_{x,\max}, \quad x = m, g, e \\ T_{x,\min} \le T_x(t) \le T_{x,\max}, \quad x = m, g, e \end{cases} \tag{14}$$

where the *max* and *min* signify the maximum and minimum values of the respective variables. The subscripts *g* and *e* designate the torque and speed pertaining to the generator and motor, respectively. For the scope of this study, we omit the consideration of the driving cycle's road conditions and the influence of temperature on battery characteristics. In the ensuing section, the DDPG algorithm is elucidated as the method employed to devise a human-like EMS by incorporating expert knowledge.

## 3. DRL Algorithm and Expert Knowledge

This section endeavors to establish a DRL framework for addressing the energy management challenge in series-parallel powertrains. Specifically, we shed light on the DDPG algorithm employed in this context. To augment the effectiveness of the DDPG algorithm, we incorporate the optimal control strategy derived from DP as expert knowledge. This incorporation serves to narrow down the search space for control actions. Additionally, we outline the procedure of gathering a real-world driving dataset from proficient drivers. Within this dataset, we detail the acquisition process of pertinent operational behaviors. These behaviors, exhibited during distinct driving cycles, are then leveraged as valuable expert knowledge for training a human-like energy management policy.

### *3.1. Dynamic Programming Method*

In line with Bellman's principle of optimality, DP allows for the acquisition of optimal global controls in multi-step horizon optimization control problems through an exhaustive exploration of state variables and control actions [24]. Numerous efforts have been made to employ DP in tackling energy management problems related to Hybrid Electric Vehicles (HEVs) [25-27]. Nevertheless, due to the curse of dimensionality, DP faces limitations in addressing problems with extensive search spaces. Consequently, the DP-based control strategy is frequently utilized as a benchmark for assessing alternative EMSs.

Bellman's principle of optimality indicates that for an *N*-steps optimization control problem, if $a(m)$ ($m$ = 1, 2, …, $N$) is the optimal control sequence over the whole time interval, then the truncated sequence $a(n)$ ($n = k+1, k+2, \ldots, N$) is still the optimal control sequence for time horizon from $k+1$ to $N$. For example, the cost function in Eq. (10) can be rewritten as follow:

$$J_N = \varphi(s_N) + \sum_{k=1}^{N-1} L(s_k, a_k, k) \tag{15}$$

where $\varphi$ is a restrictive function on the final value of SoC, and $L$ is named instantaneous cost function, which is the sum of fuel consumption rate and charge sustaining restraint. Then, the optimal cost function $J^*$ is minimizing or maximizing the cost function in Eq. (15) as:

$$J_N^* = \min\{\varphi(s_N) + \sum_{k=1}^{N-1} L(s_k, a_k, k)\} \tag{16}$$

To search the optimal control at each time step, the Eq. (16) can be further formulated as the recursive expression as:

$$J_{N-k}^* = \min\{J_{N-(k+1)}^* + L(s_k, a_k, k)\} \tag{17}$$

Executing Eq. (17) through a backward iteration process, the optimal control policy $[a^*_1, a^*_2, \ldots, a^*_N]$ is able to be computed. Then the related state variable $[s^*_1, s^*_2, \ldots, s^*_N]$ can be calculated by a recursive forward process.

In this study, the DP algorithm was utilized to attain the optimal control policy specifically for a distinct driving cycle. By applying the DP algorithm, we extracted optimal controls from a variety of real-world driving cycles that were collected. These resultant optimal control strategies were subsequently integrated into the DRL framework as expert knowledge. This integration served the purpose of constraining the search space for control actions within the DRL algorithm, ultimately augmenting the human-like attributes of the resultant adaptive energy management policy.

Put differently, the pertinent DRL algorithm avoids the search for optimal controls within the primary space. Instead, it acquires controls from diverse DP-based control policies, which stem from real-world, empirically collected driving cycles. This approach effectively reduces the search scope, resulting in enhanced computational efficiency and faster convergence rates. Moreover, considering the proficient handling of driving conditions by human drivers, the resultant energy management strategies facilitated by the DRL approach exhibit the capacity to emulate human-like attributes.

### *3.2. Historical Driving Experiences*

In real-world driving scenarios, skilled drivers adeptly adapt their driving strategies (reflected in power split controls within this study) based on the prevailing driving conditions. For instance, drivers proactively make braking decisions when approaching intersections with inoperable traffic lights. Furthermore, they may consistently engage the ICE within high-efficiency zones to optimize fuel economy during highway travel.

Taking inspiration from these well-versed and refined driving policies employed in the context of HEVs, a series of meticulously designed experiments were executed to amass a comprehensive historical driving dataset specific to HEVs. These experiments were conducted across various HEV models within Beijing, China [28]. The gathered dataset encompasses a range of parameters, including vehicle velocity, acceleration, travel distance, Global Positioning System (GPS) data, ICE and battery power outputs, as well as motor and generator torque and speed. Notably, the collected data is reflective of the routine usage of each HEV, accounting for scenarios like morning and evening traffic peaks. Consequently, this acquired dataset effectively captures diverse driving cycles occurring in both urban and highway environments.

Figure 3 illustrates the terminal device alongside its configured setup designed for data collection purposes. The parameters associated with the hybrid powertrain are meticulously captured through the distributed CAN bus system, while geographical information is diligently recorded via the integrated GPS module. This accumulated dataset is routinely transmitted to a designated cloud repository on a daily basis and is stored in the Excel file format. Notably, the sampling frequency of this data is set at 10Hz, which substantiates its precision as highly suitable for in-depth energy management investigations. The data collection initiative spanned a duration of 1320 days, encompassing 3885 data collection instances and covering a distance of 45384 km. From this extensive dataset, a subset has been thoughtfully chosen and subjected to meticulous preprocessing to serve as the basis for subsequent energy management research endeavors.

In the work presented by [28], an in-depth comparative and analytical assessment of the foundational attributes inherent in the gathered driving experiences was undertaken. These attributes encompassed various aspects, including the distribution of acceleration intervals, velocity intervals, and traction states, among others. The outcomes of this analysis strikingly underscored the pronounced dissimilarities existing between conventional standardized driving cycles and their real-world counterparts. As a result, this study selects several authentic real-world driving cycles as the basis for deriving energy management strategies that emulate human-like behavior. Specifically, these driving policies, particularly those governing power split controls between the battery and ICE, serve as the guiding principles for the search of controls within the DRL framework. These power-split controls are derived from DP calculations that take into account actual real-world driving cycles. By adopting this approach, not only is the search space for control actions significantly narrowed, but also the resultant human-like characteristics contribute to substantial improvements in both fuel economy and computational efficiency.

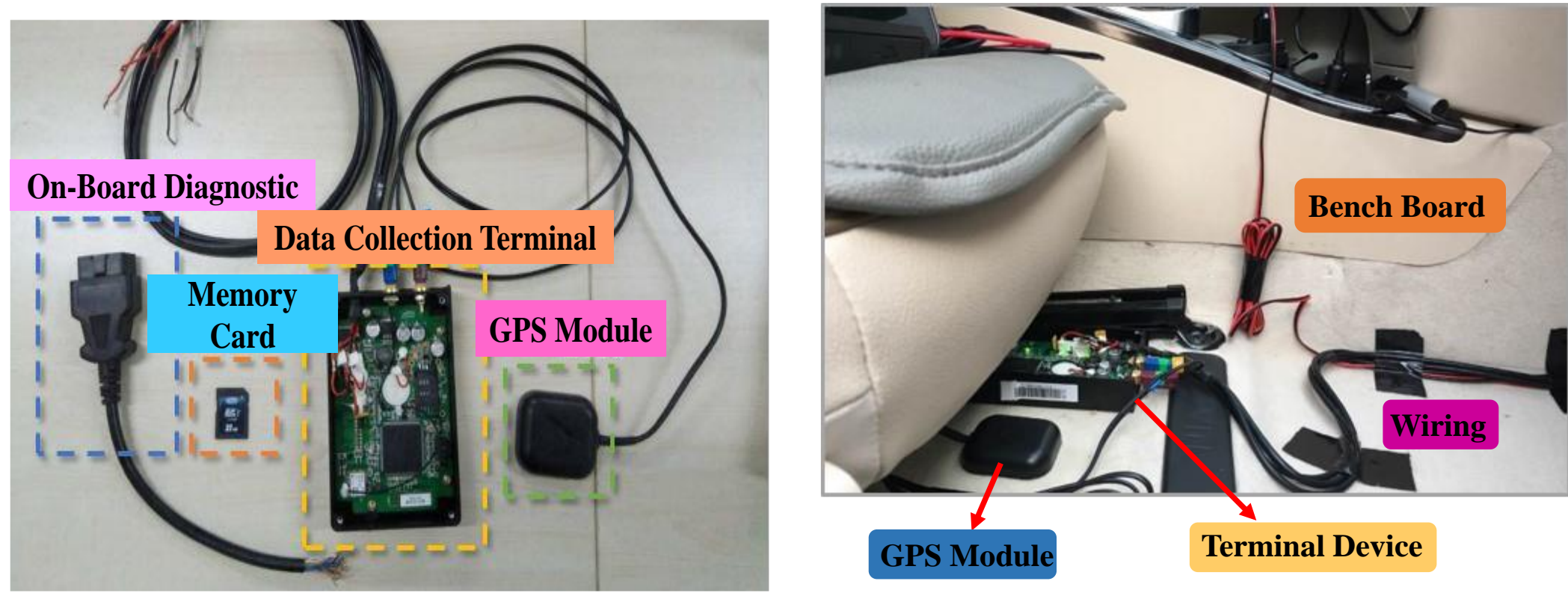

**Figure 3.** Terminal device for historical driving data collection [5].

**Table 2.** [32] Computational procedure of the modified DDPG algorithm with expert knowledge

| Modified DDPG algorithm |
| --- |
| 1. Initialize critic network $\theta^Q$ and actor network $\theta^\pi$, memory pool $K$, give initial values for $\alpha$, $\beta$, number of episodes $M$ and $\varepsilon$ |
| 2. Input DP-based control policies based on the real-world collected driving cycles (taken as expert knowledge) |
| 3. for the episode in the range [1, $M$] do |
| 4. Give initial values for three states $v_1$, $a_1$, $SoC_1$ |

5. for $t$ in the range [0, T] do
6. Choose action $a_t = \pi(s_t \mid \theta^\pi)$ according to the expert knowledge (input optimal policy) and exploration noise
7. Receive $r_t$ and $s_{t+1}$ based on current action $a_t$ and state $s_t$
8. Store transition model ($s_t$, $a_t$, $r_t$, $s_{t+1}$) in $K$
9. Sample a minibatch of transitions ($s_t$, $a_t$, $r_t$, $s_{t+1}$) from $K$ with priority experience replay
10. Set $y_t = r(s_t, a_t) + \beta Q(s_{t+1}, (a_{t+1} | \theta^\pi) \mid \theta^Q)$
11. Update the critic by minimizing the loss function: $\nabla l(\theta^Q) = E[(r + \beta Q(s_{t+1}, a_{t+1} \mid \theta^Q)) - Q(s_t, a_t \mid \theta^Q) \nabla Q(s_t, a_t \mid \theta^Q)]$
12. Update the actor policy using the sampled policy gradient: $\nabla_{\theta^\pi} L \approx E[\nabla_{\theta^\pi} Q(s_{t+1}, a_{t+1} \mid \theta^Q)] = E[\nabla_a Q(s_t, a_t \mid \theta^Q) \nabla_{\theta^\pi} \pi(s_t \mid \theta^\pi)]$
13. Store the critic and actor network for experience replay
14. end for loop
15. end for loop

### *3.3. Modified DDPG Algorithm*

In contrast to alternative machine learning approaches, RL addresses the optimization of control actions by considering the dynamic interplay between an agent and its environment [29]. The paramount strength of RL lies in its capacity for autonomous enhancement via iterative learning, often characterized as a trial-and-error exploration. This essence entails that the RL agent systematically attempts various control actions and assesses their efficacy through associated rewards. It is crucial to note that within the RL paradigm, every action carries implications not solely for immediate outcomes, but also for the ensuing rewards that follow.

Markov Decision Processes (MDPs) stand as a quintessential formalization for guiding sequential decision-making processes, wherein actions exert influence over both immediate and subsequent states and rewards [30]. As such, MDP emerges as an apt mathematical construct underpinning the realm of RL, elucidating the intricate interplay between an agent and its environment as a defined tuple $<\boldsymbol{S}, \boldsymbol{A}, \boldsymbol{P}, \boldsymbol{R}, \beta>$. Here, $\boldsymbol{S}$ denotes the set of state variables, and $\boldsymbol{A}$ encompasses the set of control actions, as detailed in Equations (11) and (12). Notably, $p \in \boldsymbol{P}$ characterizes the transition model governing state changes within the environment, while $r \in \boldsymbol{R}$ signifies the reward model instrumental in gauging action selection. Lastly, $\beta$ represents a discount factor pivotal in striking a balance between the relative significance of immediate and subsequent rewards.

The primary goal of the RL algorithm is to ascertain a control policy, denoted as $\pi$, that optimally maximizes the accumulation of rewards. In a broader context, RL algorithms are categorized into policy-based approaches, exemplified by policy gradients algorithms, and value-based approaches, represented by the Q-learning and Sarsa algorithms [29]. Within the ambit of value-based algorithms, two distinct value functions are introduced to characterize the cumulative rewards, as delineated below:

$$V^\pi(s_t) = E(\sum_{t=0}^{T} \beta r(s_t, a_t)) \tag{18}$$

$$Q^\pi(s_t, a_t) = r(s_t, a_t) + \beta \sum_{s_{t+1} \in S} p(s_{t+1} | s_t, a_t) Q(s_{t+1}, a_{t+1}) \tag{19}$$

where $V^\pi$ and $Q^\pi$ are the value functions followed by the control policy $\pi$, and $p(s_{t+1}|s_t, a_t)$ is the transition dynamic from $s_t$ to $s_{t+1}$. Then, the standard Q-learning algorithm is settled to update the action-value function $Q(s, a)$ for control action selection:

$$Q(s_t,a_t) \leftarrow Q(s_t,a_t) + \alpha[r + \beta \max_{a_{t+1}} Q(s_{t+1},a_{t+1}) - Q(s_t,a_t)] \tag{20}$$

Here, $\alpha$ represents the learning rate employed. During each discrete time step, the $\varepsilon$-greedy policy is implemented for selecting the appropriate control action. This entails that the agent would prioritize exploiting the best-known action with a probability of 1-$\varepsilon$, while also venturing into environmental exploration with a probability of $\varepsilon$. Notably, managing continuous action spaces proves challenging for Q-learning, primarily due to the vastness of the search area relative to the scope of a greedy policy. As an alternative, a solution termed DQL is introduced [31], which approximates the action-value function $Q$ utilizing a neural network (NN). This approximation is governed by a loss function articulated as follows:

$$\begin{aligned} l(\theta^Q) &= E[(Q(s_t,a_t \mid \theta^Q) - y_t)^2] \\ y_t &= r(s_t,a_t) + \beta Q(s_{t+1},a_{t+1} \mid \theta^Q) \end{aligned} \tag{21}$$

where $l(\theta^Q)$ is the loss function and $\theta^Q$ is the parameter in the neural network.

In addressing the challenge of managing the continuous spectrum of control actions, the employed approach integrates the policy gradient algorithm with a NN to compute the loss function. This integration gives rise to the Actor-Critic method, which ingeniously amalgamates the Q-learning and policy gradient algorithms [32]. Within this framework, the actor component undertakes the task of action generation through interactions with the environment, while the critic component assumes responsibility for action evaluation. As a culmination of these endeavors, the DDPG methodology is introduced, capitalizing on the advantageous features of both deep Q-learning and the Actor-Critic framework. The policy gradient for the loss function, denoted as the actor-network, is computed in the manner detailed below:

$$\begin{aligned} \nabla l(\theta^Q) = E[(r + \beta Q(s_{t+1},a_{t+1} \mid \theta^Q)) \\ - Q(s_t,a_t \mid \theta^Q)\nabla Q(s_t,a_t \mid \theta^Q)] \end{aligned} \tag{22}$$

where $\nabla$ indicates the gradient function. The critic network is approximated by the gradient via chain rule [32]:

$$\begin{aligned} \nabla_{\theta^\pi} L &\approx E[\nabla_{\theta^\pi} Q(s_{t+1},a_{t+1} \mid \theta^Q)] \\ &= E[\nabla_a Q(s_t,a_t \mid \theta^Q)\nabla_{\theta^\pi}\pi(s_t \mid \theta^\pi)] \end{aligned} \tag{23}$$

In the context of this study, the modification of the DDPG algorithm incorporates the integration of the DP-based optimal control policy or strategies gleaned from real-world data. This integration serves to streamline the search space for control actions, thus contributing to enhanced computational efficiency and overall performance. The pseudo-code outlining the adapted DDPG algorithm is presented in Table II. Within this DRL framework, key parameters are configured as follows: the learning rate $\alpha$ is set to 0.001, the discount factor $\beta$ is established at 0.95, the number of episodes amounts to 1000, the NN is allocated a memory capacity of 2000 and a batch size of 64, and $\varepsilon$ takes the form of $1*0.001^t$ (where t denotes the time step). The subsequent section of this work involves a comparative analysis of the DDPG, DP, and DQL techniques, focusing on the estimation of the proposed human-like EMS. This comparative study evaluates optimality, convergence, and adaptability through a series of carefully designed experiments.

## 4. Analyzation of Simulation Results

This section delves into an examination of the control performance exhibited by the proposed modified DDPG-based EMS (Human-like EMS). Initially, the optimality of this presented EMS is gauged through a comparison with DP, utilizing a standard driving cycle. Subsequently, an evaluation of the human-like EMS transpires on an actual real-world driving cycle1, involving a comprehensive analysis of comparative outcomes across the DDPG, DP, and DQL. Finally, the acquired human-like control policy undergoes assessment within an alternate real-world driving cycle2, shedding light on its adaptability across diverse driving scenarios within real-world environments.

### *4.1. Optimality of Modified DDPG*

In the adapted DDPG algorithm, the original range of control actions finds constraint through the integration of the DP-based control strategy, manifested as a sequential series of control actions. The proven optimality of the DP-based control sequence has been substantiated in prior studies [25]. In an effort to underscore this perspective, a comparative analysis between the human-like EMS and DP is conducted to assess its optimality across a standard driving cycle. This evaluation employs the selected cycle UDDS, showcasing two SoC trajectories stemming from these two distinct methods. The visual analysis presented in Figure 4 distinctly illustrates the close alignment of these two SoC curves, subsequently leading to the observation that the battery's output power remains largely consistent (as deduced from Equation (8), where SoC is influenced by $P_b$).

Illustrating the power allocation dynamics between the ICE and the battery, Figure 5 portrays the variation in their power outputs across the scrutinized standard driving cycle, namely the UDDS cycle. Notably, given the ICE power's characterization as the control action within this study, a discernible similarity becomes evident between the human-like EMS and its DP-based counterpart (with a few exceptions highlighted by the black circle). This congruence arises from the fact that DP is adept at establishing a globally optimal control policy through an exhaustive search process, thereby affirming the optimality inherent in the modified DDPG-enhanced EMS. In direct comparison to the conventional DDPG algorithm, the proposed control framework demonstrates its proficiency in attaining optimal control actions via an efficient learning trajectory.

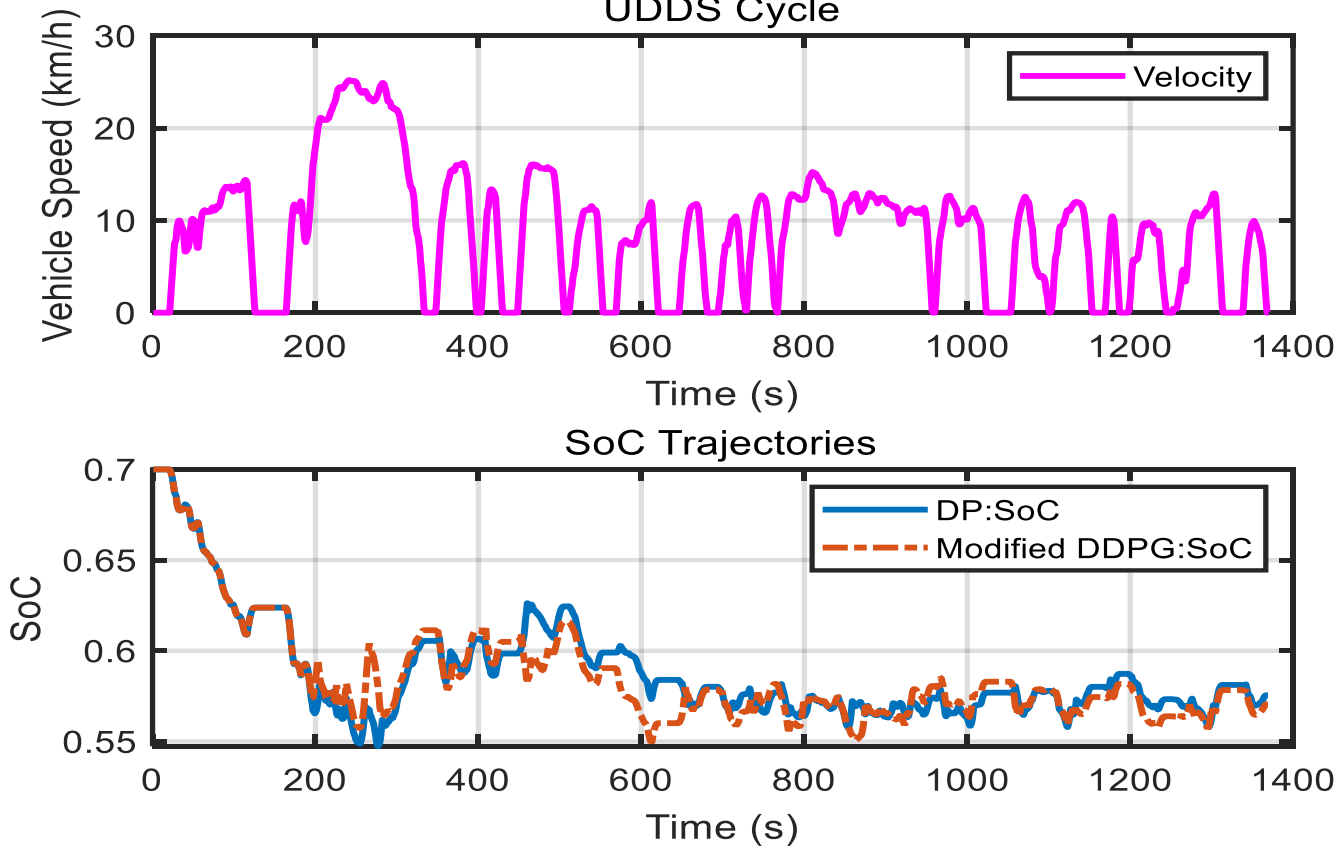

**Figure 4.** SoC curves in DP and modified DDPG on the UDDS cycle.

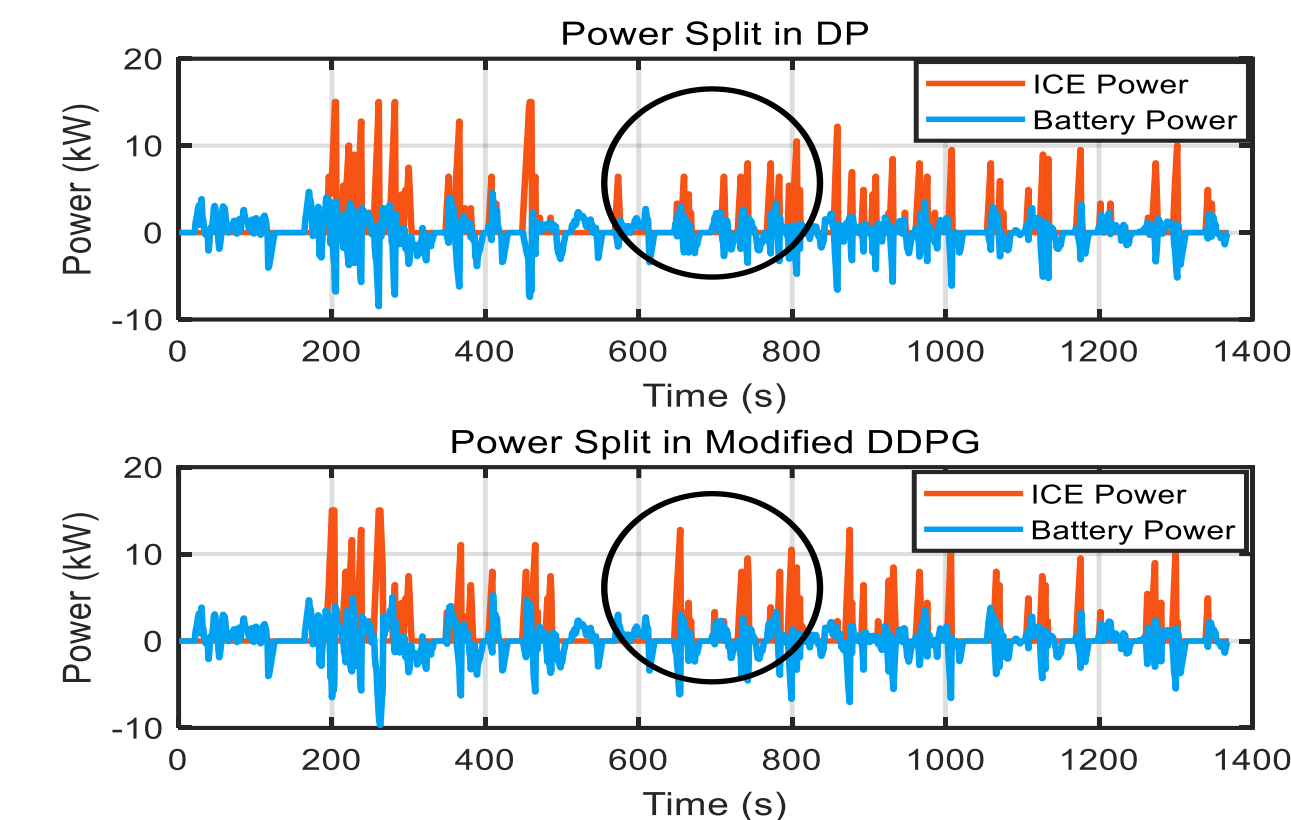

**Figure 5.** Power distribution between ICE and battery in DP and modified DDPG.

*4.2. Convergence Rate of Human-like EMS*

In this subsection, we conduct a comprehensive analysis of four distinct EMSs, derived from modified DDPG, conventional DDPG, DQL, and DP. The parameters for DQL are maintained identical to those of DDPG, ensuring an equitable basis for comparison across the three DRL methods. In the pursuit of crafting a human-like EMS for the series-parallel powertrain, a real-world driving cycle1 is employed across these four techniques, detailed in Figure 6. Moreover, Figure 7 presents the SoC curves corresponding to these four scenarios. In the context of the modified DDPG approach, the search space for control actions incorporates both the DP-based control policy and insights drawn from the strategies adopted by experienced human drivers. This design not only ensures optimality but also captures the essence of human-like characteristics. As evident, the SoC trajectories within DQL and conventional DDPG diverge from those observed in the other two instances, indicating distinct control policies. Consequently, the convergence rate and control performance of these four cases are subjected to further comparative scrutiny.

Figure 8 illustrates the total rewards across different training episodes for various DRL techniques (with a total of 1000 training episodes). Given that the reward serves as a representation of the immediate cost function in Equation (10), it becomes evident that the proposed human-like EMS is capable of achieving superior fuel economy within an equivalent number of training episodes. This outcome stems from the divergent search spaces characterizing control strategies across these instances. Consequently, it can be deduced that the modified DDPG algorithm exerts a favorable impact on fuel economy enhancement. The corresponding fuel consumption and training time for these three DRL-based EMSs are tabulated in Table III. Acknowledging that the final SoC values are not uniform, the results within Table III are standardized to an equivalent fuel consumption, effectively mitigating the influence of disparities in final SoC values [33]. It is visually evident that the modified DDPG, representing the relevant human-like EMS, consistently exhibits the most robust control performance, encompassing fuel economy and computational efficiency. As the human-like EMS framework finds application in real-world contexts, the following subsection engages in a comprehensive discussion concerning its adaptability.

**Table 3.** Equivalent Fuel Consumption for Different DRL Methods

| Techniques# | Fuel Economy* | Training Time (hours) |
|---|---|---|
| DQL | 108.9 | 10.72 |
| Conventional DDPG | 148.87 | 28.76 |
| Modified DDPG | 208.15 | 8.24 |

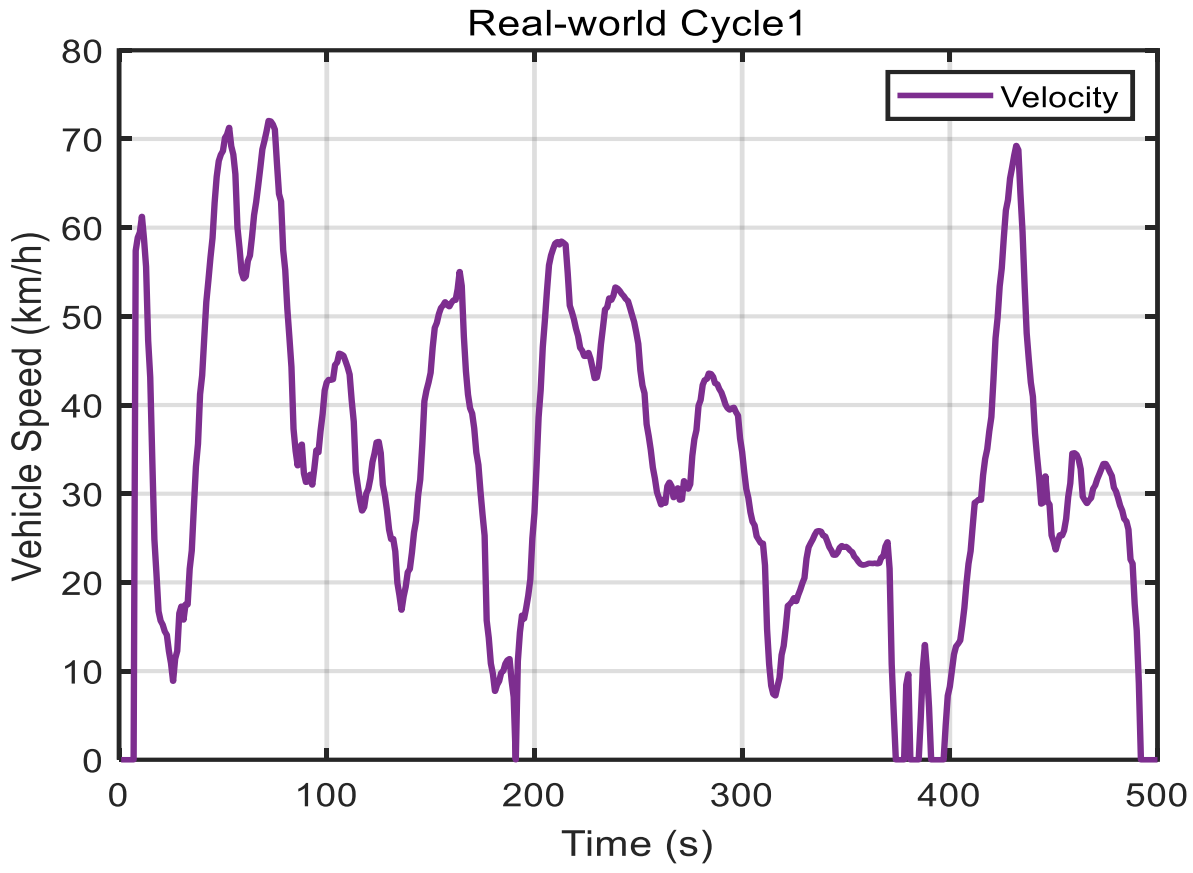

**Figure 6.** Collected real-world cycle1 for human-like EMS generation.

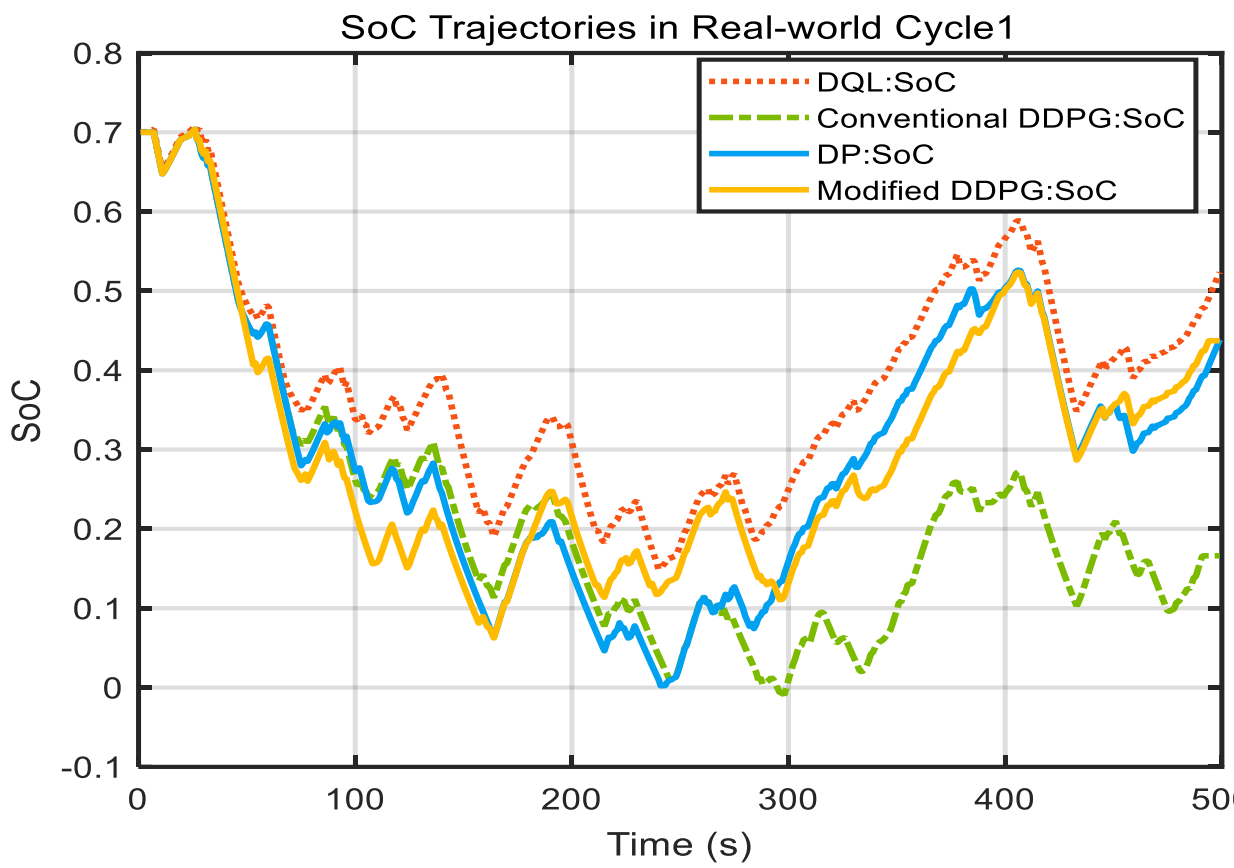

**Figure 7.** SoC variations of four compared methods on real-world driving cycle1.

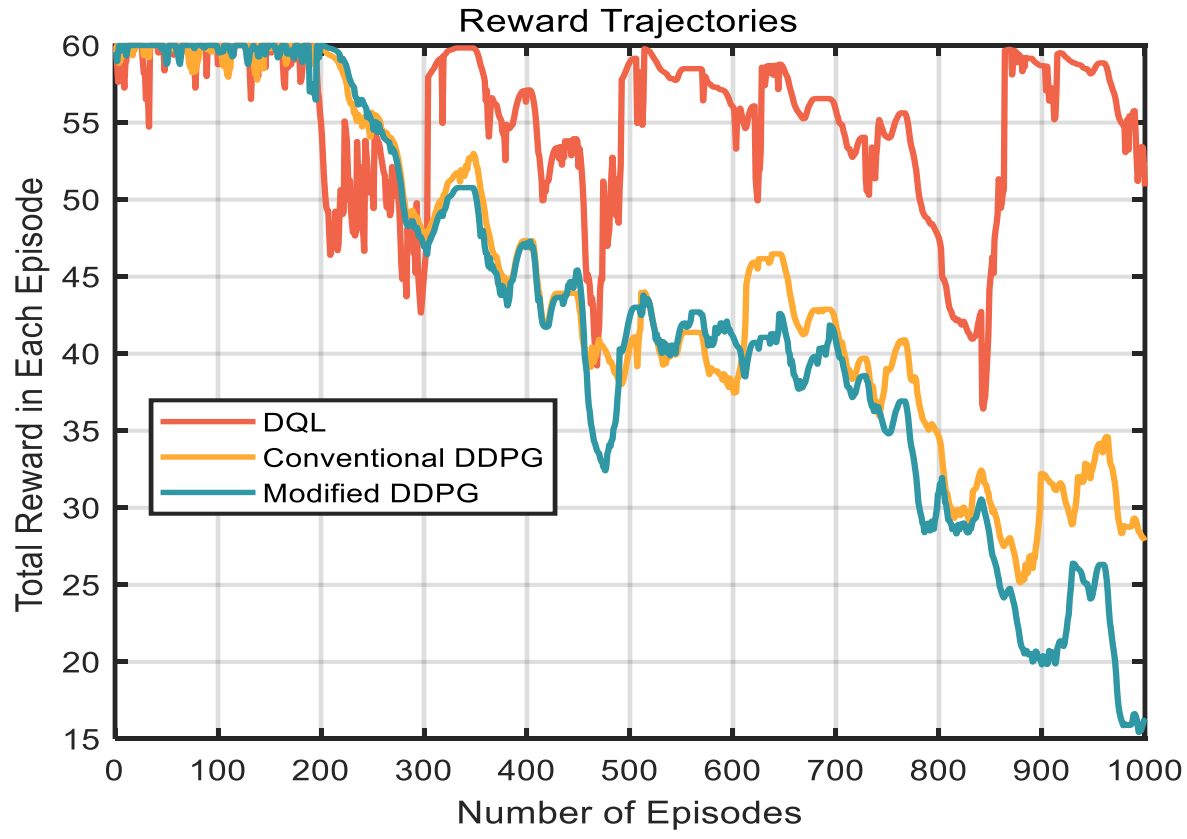

**Figure 8.** A total reward of each episode in different DRL approaches.

*4.2. Adaptability of Human-like EMS*

To provide a more comprehensive understanding of the robustness inherent in the developed human-like control policy, we subject three distinct DRL-based control strategies to evaluation within an additional real-world driving cycle, as portrayed in Figure 9. Notably, the driving cycle presented in Figure 9 remains excluded from the training process. This signifies that the critic and actor networks, originally derived from real-world driving cycle1 (as detailed in Figure 6), are subsequently applied to the distinct driving cycle2 featured in Figure 9. Through this simulation-based experimentation, insights are gleaned to ascertain the adaptability of the EMS outlined in Section IV.B to this new driving cycle. Concurrently, the pertinent SoC trajectories are visually depicted across these three scenarios within Figure 9. Notably, the diverse patterns exhibited by these SoC trajectories allude to differences in the attained control actions. This, in turn, indicates a variance in both the output power and operational points of the ICE within these respective EMSs.

Figure 10 provides an illustration of the operating range of the ICE within the context of these three DRL approaches, following an equivalent number of training episodes. Notably, the graph highlights the high-efficiency region. In this regard, the modified DDPG approach demonstrates its capacity to consistently guide the ICE towards sustained operation within this high-efficiency domain. This strategic orientation yields a noteworthy outcome - the ICE is effectively enabled to accomplish the same driving cycle while consuming notably less fuel. This achievement is attributed to the utilization of real-world driving experiences, which contribute to judicious power export from both the ICE and the battery. Consequently, the proposed human-like EMS emerges as superior in terms of fuel economy relative to alternative methodologies, thereby attesting to the adaptability embedded within the presented control framework. Furthermore, the potential for real-time application of this human-like EMS is explored through the lens of transfer learning, offering a promising avenue for future research discussions.

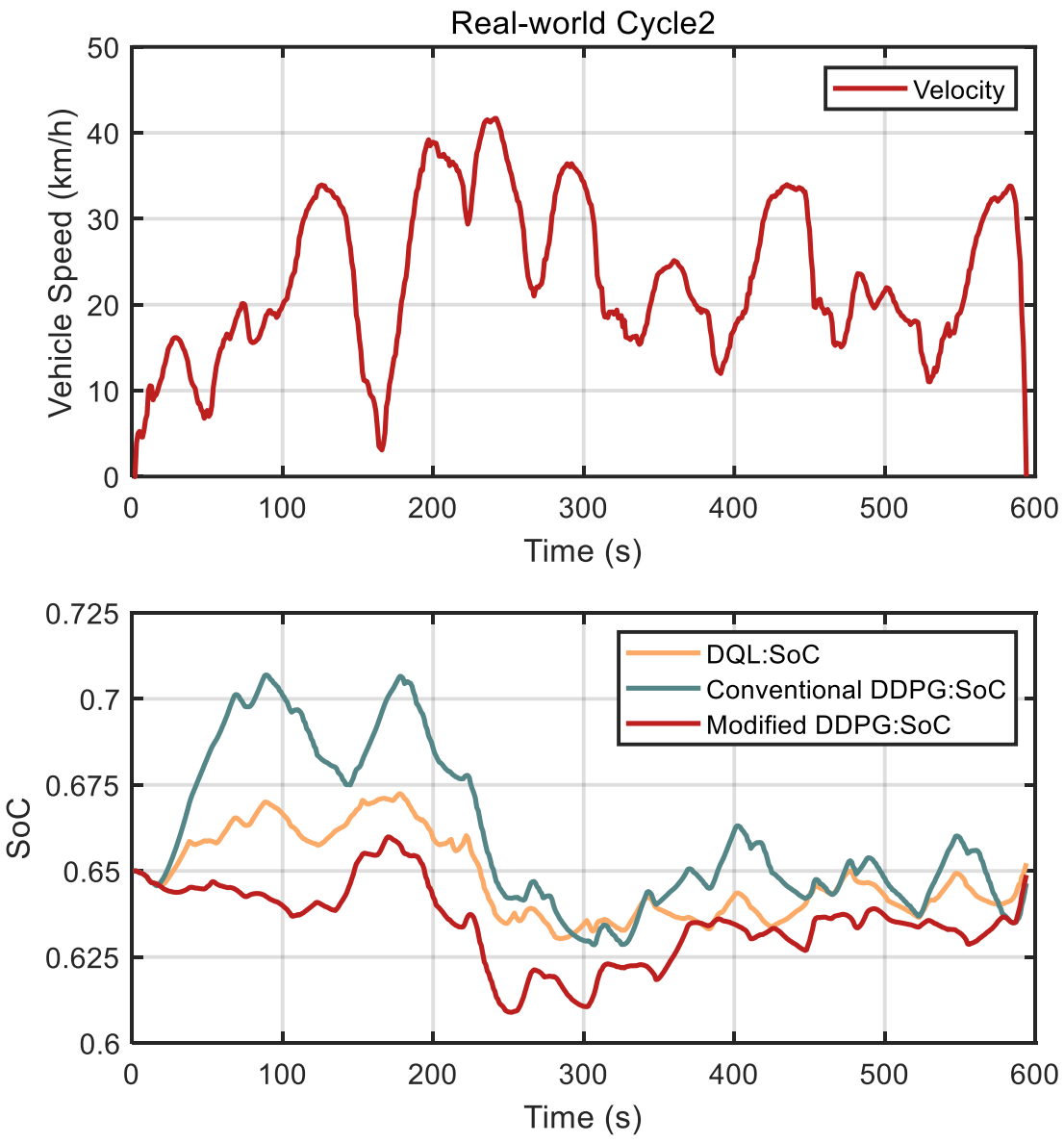

**Figure 9.** SoC trajectories of different DRL approaches on a new driving cycle.

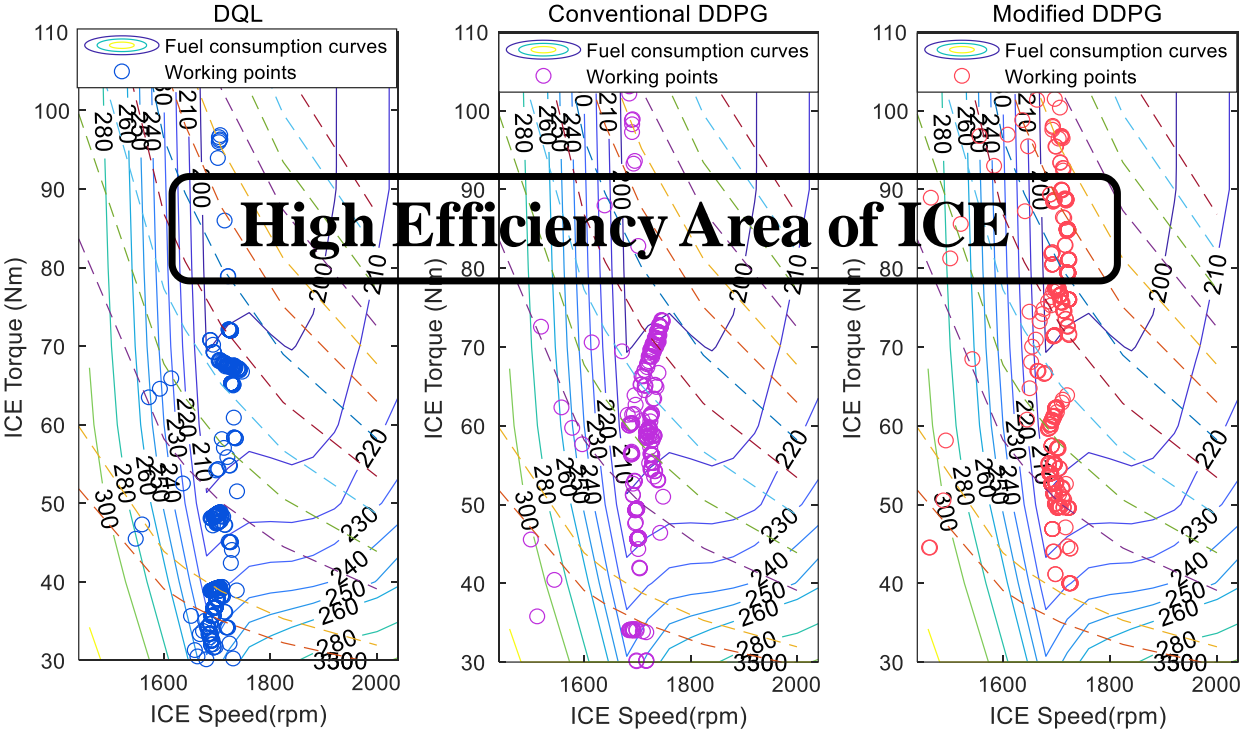

**Figure 10.** Working points of ICE in different DRL methods on a new cycle.

## 5. Conclusion

This study is centered on expediting the training process of the conventional DDPG method. This acceleration is achieved by incorporating the DP-based optimal control policy into the DRL framework, thus enhancing the optimality of the proposed EMS. Additionally, leveraging the observation that experienced human drivers adeptly adjust power distribution in alignment with real-world driving conditions, a real-world driving dataset is compiled. A subset of this dataset is thoughtfully chosen to construct a human-like control strategy. The harnessed power distribution between the ICE and the battery serves as expert knowledge, effectively guiding control choices within the DDPG algorithm. This operational paradigm empowers the intelligent agent to acquire the skill of maneuvering the hybrid powertrain akin to a human driver. Rigorous experimental design substantiates the optimality, convergence rate, and adaptability ingrained within this established human-like EMS. Notably, the findings unveil that compared to DQL and conventional DDPG, the proposed approach boosts the achieved distance by 99.25 MPG and 59.28 MPG while simultaneously trimming training time by 23.13% and 71.35%, respectively. This manifests in a superior control performance relative to DQL and conventional DDPG, with a convergence closer to DP.

Future works might concentrate on amalgamating the modified DDPG technique with transfer learning to forge an online energy management system tailored for hybrid vehicles. Given the capacity of the trained NN to exhibit generalizability across diverse real-world driving cycles, the corresponding EMS holds promise for real-time application within practical contexts. Furthermore, an extension to this research trajectory could involve the establishment of a hardware-in-the-loop (HIL) experiment, poised to showcase the practical manifestations of this envisaged direction.